\def\pmb#1{\setbox0=\hbox{#1}%
 \kern-.025em\copy0\kern-\wd0
 \kern.05em\copy0\kern-\wd0
 \kern-.025em\raise.0433em\box0}
\documentstyle[version2,preprint,aps]{revtex} 
\begin{document}
\draft 		

\begin{title}
Charge transfer fluctuation, $d-$wave superconductivity, and the\\
$B_{1g}$ Raman phonon in the Cuprates: A detailed analysis
\end{title}

\author{T. P. Devereaux$^{1}$, A. Virosztek$^{2,3}$ and A. Zawadowski$^{3,2}$
}

\begin{instit}
$^{1}$Department of Physics,
University of California,
Davis, CA 95616
\end{instit}
\begin{instit}
$^{2}$Research Institute for Solid State Physics,
POB 49,
H-1525 Budapest,
Hungary
\end{instit}
\begin{instit}
$^{3}$Institute of Physics,
Technical University of Budapest,
H-1521 Budapest,
Hungary
\end{instit}

\begin{abstract}
The Raman spectrum of
the $B_{1g}$ phonon in the superconducting
cuprate materials is investigated theoretically in
detail in both the normal and superconducting phases, and is contrasted
with that of the $A_{1g}$ phonon. A
mechanism involving the charge transfer fluctuation between the two oxygen
ions in the CuO$_2$ plane coupled to the crystal field perpendicular
to the plane is discussed and the resulting electron-phonon coupling
is evaluated. Depending on the symmetry of the phonon the weight of
different parts of the Fermi surface in the coupling is different. This
provides the opportunity to obtain information on the superconducting
gap function at certain parts of the Fermi surface.
The lineshape of the phonon is then analyzed in
detail both in the normal and superconducting states. The Fano
lineshape is calculated in the normal state and the
change of the linewidth with temperature below T$_{c}$ is
investigated for a $d_{x^{2}-y^{2}}$ pairing symmetry. Excellent
agreement is obtained for the $B_{1g}$ phonon lineshape in
YBa$_{2}$Cu$_{3}$O$_{7}$. These experiments, however, can not distinguish
between $d_{x^{2}-y^{2}}$ and a highly anisotropic $s$-wave pairing.
\end{abstract}
\receipt{\today}
\pacs{PACS numbers: 74.20.-z, 74.20.Fg, 74.25.Gz, 74.25.Kc, 74.72.-h}
\narrowtext

\section{Introduction}

As a result of the recent dispute as to whether the superconducting gap
in the high-T$_c$ materials is $s-$ or $d-$type all experiments which
can contribute to resolve that question are of special importance\cite{PAS}.
It was shown recently\cite{Dev} that the study of the polarization
dependence of the electronic Raman scattering can address these questions
and can distinguish between the different orientations ($d_{x^2-y^2}$ or
$d_{xy}$) of the gap as well. This evidence can be made more complete
by including the phonons and the effect of the electron-phonon coupling,
which is the subject of the present paper\cite{tpd}.

The analysis of the lineshape of optical phonons in the high T$_{c}$
superconductors has been the focus of a large body of
investigation\cite{Cooper,rev}.
Of the many Raman active modes in tetragonal superconductors, all
the modes that have been observed in the cuprate materials using in-plane
polarizations transform
according the $A_{1g}$ symmetry except one, which transforms according
to $B_{1g}$ symmetry\cite{rev}. Since this phonon obeys its own selection
rules, the
frequency of the mode can be unambiguously assigned, which is in contrast
with the early confusion of the assignments of the separate $A_{1g}$
modes.

In this paper the phonons involving the perpendicular vibrations of the
in-plane oxygens are investigated. The displacements of the two oxygens
O(2) and O(3) are in phase for $A_{1g}$ symmetry and out of phase for
$B_{1g}$ as is shown on Fig. (1) for the case of YBa$_2$Cu$_3$O$_y$.
The coupling of these phonons to the
electrons in the plane has an interesting feature, namely a completely
flat CuO$_2$ plane taken out of the surrounding atomic environment has
the mirror symmetry through the plane. Thus the linear coupling between
these phonons and the in-plane electrons is absent. In a crystal however,
as it has first been pointed out by Bari\v si\' c {\it et.al.}\cite{Bar},
that symmetry can be broken since there is an electric field perpendicular
to the plane due to the surrounding ions forming an asymmetric
environment. This is in contrast to the La compound, where only the small
tilting of the octahedra breaks the symmetry\cite{Bar2}. The
perpendicular electric field in the 1:2:3 material
can be responsible for the buckling of the CuO$_2$ plane, i.e. the slight
separation of the planes formed by the Cu and O ions. This small
distortion however cannot be responsible for the large electron-phonon
coupling to be discussed here.

Since the long-range charge transfer fluctuations between unit
cells are screened, only intracell charge transfers must be considered.
Let us denote the charge transfer fluctuations at O(2) and O(3) by
$\delta\rho_x$ and $\delta\rho_y$ (see Fig. 1.) respectively. A charge
transfer fluctuation where $\delta\rho_x-\delta\rho_y\not= 0$ is
obviously coupled to the electric field and the crystal deforms in the
form of the $B_{1g}$ phonon, similarly $\delta\rho_x +\delta\rho_y\not= 0$
is coupled to the $A_{1g}$ phonon. The fluctuations of the in-plane
charges are either due to the transfer (i) between the in-plane
oxygens and other ions, or (ii) between the two in-plane oxygens. The
fluctuations of the $A_{1g}$ symmetry $\delta\rho_x =\delta\rho_y$ can
be a consequence of the transfers between the in-plane copper and oxygens.
For $B_{1g}$ symmetry both (i) and (ii) may be realized.

Considering the first possibility\cite{MZ}, the charge transfer in 1:2:3
between
the bridging oxygen O(4) and the O(3) is a good candidate.
As long as the non-bonding
orbital of O(4) parallel to the CuO chain has a partial electronic density
of states near the Fermi surface, the transfer breaks the $x-y$ symmetry.
That mechanism, which is sensitive
to the position of the electronic partial density of states of the
bridging oxygen relative to the Fermi energy, has been worked out in
detail\cite{MZ}, but the recently observed Raman spectra are not consistent
with
the predicted $d_{xy}$ superconducting gap\cite{Dev}.

The second possibility does not depend on such details of the electronic
band structure, since only the conduction band plays a role.
The charge transfer fluctuation $\delta\rho_x +
\delta\rho_y =0$ between O(2) and O(3) involves especially those parts of the
cylindrical Fermi surface which are close to the $k_x$ and $k_y$ axis.
Thus the influence of the electron-phonon coupling on the electronic
Raman continuum depends strongly on whether a large superconducting gap
opens at that part of the Fermi surface. More precisely, the electronic
contribution can be characterized by the azimuthal quantum number $m$
on the cylindrical Fermi surface\cite{tpd,Klein}, and $m=\pm 2$ corresponds to
$B_{1g}$
symmetry, while out of the two ($m=0$, $m=\pm 4$) channels of $A_{1g}$
symmetry only $m=\pm 4$ is relevant since the $m=0$ mode describes charge
fluctuations between cells, which are screened by the long-range Coulomb
interaction. The different Fermi surface areas probed by Raman
scattering in the $A_{1g}$ and $B_{1g}$ geometry are illustrated in
Fig. (2). ($A_{1g}$ and $B_{1g}$ transforms like $\cos(4\phi)$ and
$\cos(2\phi)$ respectively).
Thus assuming a $d_{x^2-y^2}$ type superconducting gap the $B_{1g}$
mode probes the areas with the largest gap, while the $A_{1g}$ averages
the areas of both the largest gap and the nodes. The most appropriate tool
in the Raman experiments to probe the electron-phonon coupling is to
study the Fano interference which is formed due to the simultaneous
scattering of light on the phonons and the electronic continuum
(see eg. Refs. \cite{Cooper} and \cite{MZ}). The fit
of the Fano lineshape provides information not only on the coupling but
also on the electronic continuum influenced by the opening of the
superconducting gap. The appearance of a sharper gap in the $B_{1g}$
symmetry than in the $A_{1g}$ provides a unique identification of the gap
of $d_{x^2-y^2}$ type\cite{Dev}. It must be emphasized however, that these
experiments are not sensitive to the sign of the order parameter, therefore
can not distinguish between $d$-wave superconductivity and a very highly
anisotropic $s$ type.

The present status of the relation between experiment and theory can be
summarized as follows.
The vibration of the $B_{1g}$ mode which appears at roughly
340 wavenumbers in all cuprate materials, is connected with the
antisymmetric out-of-plane vibrations of the O(2) and O(3) ions in
the Cu-O plane\cite{rev}. The net charge transfer of this vibration in the
unit cell is zero and thus the long range Coulomb forces are incapable
of screening the charge fluctuations. Consequently, the mode has a
large cross section for scattering incoming photons and thus appears
in Raman experiments as a large sharp signal centered at frequency
shifts corresponding to the energy of the mode. Due to the strength and
unambiguous identification of the mode, the $B_{1g}$ phonon has been
lavished with attention\cite{rev}. In particular, the spectral lineshape
of the
mode has been a subject of intense scrutiny both in the normal and
superconducting state of the cuprates. In the normal state, the asymmetry
of the lineshape indicates strong mixing of the phonon with the
electronic continuum and fits to a Breit-Wigner or Fano lineshape have
been made with great success\cite{Cooper}. This has usually led to the
assertion that
the electron-phonon coupling at least for this mode is large. However
no estimates of the coupling from a microscopic theory have been presented
except for the theoretical suggestion by Bari\v si\' c and Kup\v ci\' c
\cite{Bar}.
Further, the anomalous changes of the $B_{1g}$ phonon as the material
enters into the superconducting state have also been exhaustively
documented\cite{Cooper,rev,Had}. The temperature dependence of the changes
in the $B_{1g}$
phonon's lineshape have been used to help determine the magnitude of
the energy gap of the superconductor\cite{Had}. Recently, a theory of Raman
scattering in $d-$wave superconductors has been presented\cite{Dev,tpd}
and while the
general features of the theory fit well with the experiment, no detailed
fit of the lineshape could be made without further knowledge of the
mechanism and strength of the electron-phonon coupling.

In the present paper we provide a detailed theory for the behavior of
the $B_{1g}$ phonon in both the normal and superconducting states of
the cuprate materials using the mechanism suggested by Bari\v si\' c
{\it et.al.}\cite{Bar,Bar2}. In particular, we investigate the mechanism
of electron-phonon coupling resulting from crystal field effects and
describe the Fano lineshape
in the normal state. The theory is then generalized to the
superconducting state and the lineshape is calculated in detail for a
superconductor with $d_{x^{2}-y^{2}}$ pairing.

The paper is organized as follows. In Sec. II. we develope the mechanism
which leads to first order electron-phonon coupling due to the
presence of the crystal field. Based on the 3-band model for the CuO$_2$
plane the electron-phonon coupling constant is evaluated. In Sec. III.
we apply these results in order to fit the experimental data in the
normal state. Particular attention is paid to the Fano resonance.
Section IV. is devoted to the behavior of the phonon lineshape in the
superconducting state. Here we show that the temperature dependence of
the spectrum is due to the change in the electronic response, and the
data are consistent with a $d_{x^2-y^2}$ gap. Finally our conclusions
are given in Sec. V.

\section{mechanism}

In the first part of this section we define our model for the electrons
in the CuO$_2$ plane. This allows us to introduce notations used in the
second part, where we develop a microscopic theory of
the coupling of these electrons to the
out-of-plane phonons due to the crystal field perpendicular to the plane.

\subsection{The model}

Using the notations of Ref.\cite{Zawa}
we consider the 3-band model for the CuO$_2$ plane described
by the following Hamiltonian:
\begin{equation}
H^0=\varepsilon\sum_{{\bf n},\sigma} b^{\dagger}_{{\bf n},\sigma}
b_{{\bf n},\sigma}+t\sum_{{\bf n},\pmb{$\delta$},\sigma} P_{\pmb{$\delta$}}
(b^{\dagger}_{{\bf n},\sigma} a_{{\bf n},\pmb{$\delta$},\sigma}+h.c.),
\end{equation}
where $b^{\dagger}_{{\bf n},\sigma}$ creates an electron with spin
$\sigma$ at a copper lattice site ${\bf n}$, while $a_{{\bf n},\pmb{$\delta$},
\sigma}$ annihilates an electron at one of the neighboring oxygen sites
${\bf n}+\pmb{$\delta$}/2$
determined by the unit vector $\pmb{$\delta$}$ assuming the four values,
$(\pm 1,0)$ and $(0,\pm 1)$. An oxygen atom between the two copper atoms at
sites ${\bf n}$ and ${\bf n}+\pmb{$\delta$}$ is labeled by either
$({\bf n},\pmb{$\delta$})$ or $({\bf n}+\pmb{$\delta$},-\pmb{$\delta$})$.
Moreover, $\varepsilon=E_d-E_p$ is the difference of the Cu and O site
energies, $t$ is the Cu-O hopping integral and $P_{\pmb{$\delta$}}=\pm 1$
depending on whether the orbitals (with real wavefunctions) have the same
or opposite sign at the overlap region. Assuming Cu $d_{x^2-y^2}$ and
O $p$ orbitals $P_{-\pmb{$\delta$}}=-P_{\pmb{$\delta$}}$, and we can choose
$P_{(1,0)}=1$ and $P_{(0,1)}=-1$. For simplicity, we neglect direct O-O
hopping.

The momentum representation is defined by the following formulae:
\begin{equation}
b_{{\bf n},\sigma}={1\over\sqrt{N}}\sum_{\bf k} e^{i{\bf k}a{\bf n}}
b_{{\bf k},\sigma},
\end{equation}
and
\begin{equation}
a_{{\bf n},\pmb{$\delta$},\sigma}={1\over\sqrt{N}}\sum_{\bf k}
\exp[i{\bf k}a({\bf n}+\pmb{$\delta$}/2)] a_{\alpha,{\bf k},\sigma},
\end{equation}
where $a$ is the lattice constant, $N$ is the number of Cu sites and
$\alpha$ is $x$ or $y$ for $\pmb{$\delta$}=(\pm 1,0)$ and $\pmb{$\delta$}=
(0,\pm 1)$ respectively. In this representation the Hamiltonian in Eq.(1)
decouples for different momenta as
\begin{equation}
H^0=\sum_{{\bf k},\sigma} H^0_{{\bf k},\sigma},
\end{equation}
where
\begin{equation}
H^0_{{\bf k},\sigma}=\varepsilon b^{\dagger}_{{\bf k},\sigma}b_{{\bf k},
\sigma} +\{ ib^{\dagger}_{{\bf k},\sigma}[a_{x,{\bf k},\sigma}t_x({\bf k})
-a_{y,{\bf k},\sigma}t_y({\bf k})]+h.c.\},
\end{equation}
and
\begin{equation}
t_{\alpha}({\bf k})=2t\sin(ak_{\alpha}/2).
\end{equation}

$H^0_{{\bf k},\sigma}$ is then easily diagonalized as
\begin{equation}
H^0_{{\bf k},\sigma}=\sum_{\beta} E_{\beta}({\bf k})d^{\dagger}_{\beta,
{\bf k},\sigma}d_{\beta,{\bf k},\sigma},
\end{equation}
where $\beta$ assumes the values $+$, $-$, and $0$ for the antibonding,
bonding, and nonbonding bands, respectively. The corresponding electronic
energies are $E_0({\bf k})=0$ and
\begin{equation}
E_{\pm}({\bf k})=\varepsilon/2\pm\sqrt{(\varepsilon/2)^2+\Omega^2({\bf k})},
\end{equation}
with
\begin{equation}
\Omega^2({\bf k})=t^2_x({\bf k})+t^2_y({\bf k}).
\end{equation}
The transformation between the original ($a,b$) and the new ($d$) electronic
operators is described by
\begin{equation}
\left ( \begin{array}{c} b \\ a_x \\ a_y \end{array}\right ) =
({\bf e}_+,{\bf e}_0,{\bf e}_-)
\left ( \begin{array}{c} d_+ \\ d_0 \\ d_- \end{array}\right ),
\end{equation}
where the column vectors of the transformation matrix are given by
\begin{equation}
{\bf e}_0={1\over \Omega}
\left ( \begin{array}{c} 0 \\ t_y \\ t_x \end{array}\right ),
\end{equation}
and
\begin{equation}
{\bf e}_{\pm}={1\over \sqrt{E^2_{\pm}+\Omega^2}}
\left ( \begin{array}{c} E_{\pm} \\ -it_x \\ it_y \end{array}\right ).
\end{equation}
In the last three equations we have dropped the ${\bf k},\sigma$ indices
for clarity.

In the physically relevant situation for the CuO$_2$ plane the upper band
is close to half filled. Since the three bands do not overlap in the
present model, we will consider only a reduced one-band Hamiltonian
describing the upper band and neglect the bonding and nonbonding
bands. Then the reduced Hamiltonian is given by
\begin{equation}
H^0_{red}=\sum_{{\bf k},\sigma} E({\bf k})d^{\dagger}_{{\bf k},\sigma}
d_{{\bf k},\sigma},
\end{equation}
with
\begin{equation}
E({\bf k})=\varepsilon/2+\sqrt{(\varepsilon/2)^2+\Omega^2({\bf k})}.
\end{equation}
Furthermore, the expressions of the transformation from $d$ operators
to $a$ and $b$ operators reduce to
\begin{mathletters}
\begin{equation}
b_{{\bf k},\sigma}=\phi_b({\bf k})d_{{\bf k},\sigma},
\end{equation}
\begin{equation}
a_{x,{\bf k},\sigma}=\phi_x({\bf k})d_{{\bf k},\sigma},
\end{equation}
\begin{equation}
a_{y,{\bf k},\sigma}=\phi_y({\bf k})d_{{\bf k},\sigma},
\end{equation}
\end{mathletters}
where
\begin{mathletters}
\begin{equation}
\phi_b({\bf k})=E({\bf k})[E^2({\bf k})+\Omega^2({\bf k})]^{-1/2},
\end{equation}
\begin{equation}
\phi_x({\bf k})=-it_x({\bf k})[E^2({\bf k})+\Omega^2({\bf k})]^{-1/2},
\end{equation}
\begin{equation}
\phi_y({\bf k})=it_y({\bf k})[E^2({\bf k})+\Omega^2({\bf k})]^{-1/2}.
\end{equation}
\end{mathletters}
As we will see in the next subsection, this simple model contains the
most important ingredients in order to describe the electron-phonon coupling.

\subsection{Electron-phonon coupling}

As we have discussed already, the electrons in the CuO$_2$ plane do not
couple in first order to the phonon modes with displacement vectors
perpendicular to the plane, because in that case
the hopping integrals change only in second order in the displacements.
However, if there is an electric field perpendicular to the plane, first
order electron-phonon coupling is generated. In our case this field
originates from the asymmetric environment around the CuO$_2$ plane
(see Fig.1.), and is also responsible for the buckling of the plane with
restoring force provided by the covalent bonds.

Let us consider the effect of a lattice periodic crystal field
${\bf E}({\bf r})=-\nabla \phi_{ext}({\bf r})$ on the electrons
in the plane. The electron density at each (displaced) site couples to
the external field via the Hamiltonian
\begin{eqnarray}
 &H^{\prime}=-e\sum_{{\bf n},\sigma} \{ b^{\dagger}_{{\bf n},\sigma}
b_{{\bf n},\sigma}\phi_{ext}[a{\bf n}+{\bf u}_b(a{\bf n})]\nonumber \\
+&a^{\dagger}_{{\bf n},{\bf x},\sigma}a_{{\bf n},{\bf x},\sigma}
\phi_{ext}[a{\bf n}+a{\bf x}/2+{\bf u}_x(a{\bf n})]\nonumber \\
+&a^{\dagger}_{{\bf n},{\bf y},\sigma}a_{{\bf n},{\bf y},\sigma}
\phi_{ext}[a{\bf n}+a{\bf y}/2+{\bf u}_y(a{\bf n})]\},
\end{eqnarray}
where ${\bf u}_b(a{\bf n})$, ${\bf u}_x(a{\bf n})$, and ${\bf u}_y(a{\bf n})$
are displacement vectors of the Cu, O(2), and O(3) in the unit cell at the
lattice site ${\bf n}$, $e$ is the electron charge, and ${\bf x}$ and
${\bf y}$ are unit vectors in the corresponding directions. Expansion in the
displacements up to first order leads to
\begin{equation}
H^{\prime}=H^{\prime}_{site}+H_{el-ph}+...,
\end{equation}
where $H^{\prime}_{site}$ renormalizes the copper-oxygen site energy
difference $\varepsilon$ only, while the term linear in ${\bf u}$ generates
an electron-phonon interaction
\begin{eqnarray}
 &H_{el-ph}=e\sum_{{\bf n},\sigma}\{ {\bf E}_b {\bf u}_b(a{\bf n})
b^{\dagger}_{{\bf n},\sigma}b_{{\bf n},\sigma}\nonumber \\
+&{\bf E}_x {\bf u}_x(a{\bf n})a^{\dagger}_{{\bf n},{\bf x},\sigma}
a_{{\bf n},{\bf x},\sigma}\nonumber \\
+&{\bf E}_y {\bf u}_y(a{\bf n})a^{\dagger}_{{\bf n},{\bf y},\sigma}
a_{{\bf n},{\bf y},\sigma}\},
\end{eqnarray}
where ${\bf E}_b$, ${\bf E}_x$, and ${\bf E}_y$ are the electric fields
at the Cu, O(2), and O(3) sites respectively, which are parallel to the
$z$-axis due to symmetry. In the absence of chains ${\bf E}_x={\bf E}_y$
would hold.

$H_{el-ph}$ can be written in momentum representation with the help of
Eqs.(2-3), and Eq.(15) allows us to express the interaction
of phonons and electrons of the reduced Hamiltonian Eq.(13) in the usual
form
\begin{equation}
H_{el-ph}={1\over\sqrt{N}}\sum_{{\bf q},\lambda}\sum_{{\bf k},\sigma}
g_{\lambda}({\bf k},{\bf q})d^{\dagger}_{{\bf k}+{\bf q},\sigma}
d_{{\bf k},\sigma}[c_{\lambda}({\bf q})+c^{\dagger}_{\lambda}(-{\bf q})].
\end{equation}
Here $c_{\lambda}({\bf q})$ annihilates a phonon mode $\lambda$ with wavevector
${\bf q}$ and $g_{\lambda}({\bf k},{\bf q})$ is the coupling constant of that
mode to an electron of wavevector ${\bf k}$. Based on our microscopic model
the coupling constant can easily be evaluated using standard procedures of
the quantum theory of phonons\cite{AM}.

In the context of Raman scattering we are interested in ${\bf q}=0$ optical
phonons. In case of the $B_{1g}$ phonon the Cu displacement is zero, while
the O(2) and O(3) atoms have equal and opposite displacements (Fig.1.b.).
The corresponding coupling constant is evaluated as
\begin{equation}
g_{B_{1g}}({\bf k},{\bf q}=0)=e\sqrt{\hbar\over 2M_o\omega_{B_{1g}}}
{1\over\sqrt{2}}[({\bf E}_x)_z|\phi_x({\bf k})|^2-({\bf E}_y)_z
|\phi_y({\bf k})|^2],
\end{equation}
where $M_o$ is the oxygen mass and $\omega_{B_{1g}}$ is the phonon frequency.
In the following we make the approximation $({\bf E}_x)_z=({\bf E}_y)_z=E_z$.
Due to the opposite displacements of the oxygens, the {\it difference} of the
$\phi_{\alpha}$ functions (given by Eq.(16)) determines the
${\bf k}$ dependence of the $B_{1g}$ coupling.
Since $\Omega({\bf k})=const.$ on the Fermi
surface (see Eq.(14)), this coupling constant is proportional to
$t^2_x({\bf k})-t^2_y({\bf k})\propto \cos(ak_x)-\cos(ak_y)$, which is
clearly of $B_{1g}$ character. In fact, this is exactly the second order
Fermi surface harmonic\cite{Allen} of our model band structure (Eq.(14))
transforming according to the $B_{1g}$ symmetry. Therefore we can write
the coupling in the form
\begin{equation}
g_{B_{1g}}({\bf k},{\bf q}=0)=g_{B_{1g}}\phi^{m=2}_{B_{1g}}({\bf k}),
\end{equation}
where
\begin{equation}
\phi^{m=2}_{B_{1g}}({\bf k})={1\over\sqrt{N_{B_{1g}}(E_F)}}[\cos(ak_x)-
\cos(ak_y)]
\end{equation}
is the normalized second order Fermi surface harmonic of $B_{1g}$ symmetry with
the normalization constant at the Fermi energy $E_F$ given by\cite{Allen}
\begin{equation}
N_{B_{1g}}(E_F)={\int d^2 k \delta [E_F-E({\bf k})][\cos(ak_x)-
\cos(ak_y)]^2\over\int d^2 k \delta [E_F-E({\bf k})]}.
\end{equation}
All information about the strength of the coupling is now compressed in
the expansion coefficient
\begin{equation}
g_{B_{1g}}=-eE_z\sqrt{\hbar\over 2M_o\omega_{B_{1g}}}{2t^2\over E_F
(2E_F-\varepsilon)}\sqrt{N_{B_{1g}}(E_F)\over 2}.
\end{equation}

In case of the $A_{1g}$ phonon the oxygen displacements are the same both
in magnitude and in direction (Fig.1.a) and the displacement of the
copper is negligible due to the relative rigidity of its vertical bonds.
The same procedure employed for the $B_{1g}$ phonon yields the following
coupling constant for the $A_{1g}$ phonon
\begin{equation}
g_{A_{1g}}({\bf k},{\bf q}=0)=eE_z\sqrt{\hbar\over 2M_o\omega_{A_{1g}}}
{1\over\sqrt{2}}[|\phi_x({\bf k})|^2+|\phi_y({\bf k})|^2].
\end{equation}
In this case however, the coupling depends on ${\bf k}$ only through
the energy $E({\bf k})$, i.e. it is constant on the Fermi surface
for any band filling. Therefore within the present model the $A_{1g}$
phonon couples to homogeneous density fluctuations only, and since these
fluctuations are suppressed by the long range Coulomb interaction, the
resulting coupling is vanishingly small. This conclusion remains valid
if we allow for a finite Cu displacement as well\cite{Cud}, but does not
necessarily hold if e.g. O-O hopping is included.

At the end of this section we wish to evaluate the coupling in Eq.(25).
For a half filled band the Fermi energy $E_F=\varepsilon/2+
\sqrt{(\varepsilon/2)^2+(2t)^2}$, and the normalization constant
$N_{B_{1g}}(E_F)=4$.
It is reasonable to suppose that $\varepsilon/2\ll 2t$, and in this limit
we only need the electric field $E_z$ at the oxygen site for a numerical
value of $g_{B_{1g}}$. In order to estimate this field we suppose that
the charges on the planes of the unit cell of YBa$_2$Cu$_3$O$_7$ are
evenly distributed, and the charges (per unit cell, in units of $e$) for
the planes Y, Ba, and CuO$_2$ are +3, +2, and -2 respectively, while the
remaining (chain) region has -3 electron charge to insure neutrality of the
unit cell. Each plane produces an electric field at the oxygen site we are
concerned with independent of its distance. Since all planes in the sample
contribute, we consider ever larger environments of the CuO$_2$ plane in
question. We calculate the field produced by the two neighboring (Y and Ba)
planes first, then include the next nearest neighbors etc. The series of
values for $E_z$ is of course not convergent, but has a period of the unit
cell. Although a more accurate calculation applying the Ewald summation
is desirable\cite{Bar2},
for an estimate we use the average value in this series, which
yields $eE_z=-2\pi e^2/a^2=-6.1$eV/\AA. According to Eq.(25) this crystal
field generates a coupling $g_{B_{1g}}=0.12$eV. The relevant dimensionless
coupling $\lambda_{B_{1g}}=N_{F}g^2_{B_{1g}}/\hbar\omega_{B_{1g}}=0.078$ if
we use a total density of states at the Fermi level
$N_{F}=0.22$/eV corresponding to a bandwidth of 9eV.
We note here that due to the presence of chains the tetragonal symmetry
of the CuO$_2$ plane is weakly broken, and $({\bf E}_x)_z\not= ({\bf E}_y)_z$.
This leads to the slight mixing of the $A_{1g}$ and $B_{1g}$ modes.

\section{normal state}

In this section we utilize our calculations for the coupling
constant to discuss the resulting spectra of the cuprates in the
normal state, while in the next section we discuss the changes in
the spectra due to superconductivity.

The inelastic scattering of photons from a metal can either be
caused by collisions with phonons or via the creation of
electron-hole pairs. As well, the phonons interact with the
electronic continuum and have a dynamical effect on the way
photons are scattered. The total light scattering cross section
resulting from these contributions is depicted by the Feynman
diagrams in Fig. 3. The coupling constant $\gamma$ describes the
electron-photon coupling, (i.e., the Raman vertex
for particle-hole creation due to the vector potential of the
incoming light). In the limit of small momentum transfer and
frequency transfers smaller than the optical band gap, the Raman
vertex is
simply related to the curvature of the band dispersion $\epsilon({\bf k})$
and the incident $({\bf \hat e^{I}})$ and scattered $({\bf \hat e^{S}})$
polarization light vectors via
\begin{equation}
\gamma({\bf k})={m\over{\hbar^{2}}}\sum_{\alpha,\beta}e^{I}_{\alpha}
{\partial^{2}\epsilon({\bf k})\over{\partial k_{\alpha}\partial k_{\beta}}}
e^{S}_{\beta},
\end{equation}
which can be expanded in terms of Fermi surface harmonics $\phi^{m}({\bf k})$,
$\gamma({\bf k})=\sum_{m}\gamma_{m}\phi^{m}({\bf k})$.
The remaining vertices we denote by
$g_{p-p}$ for the photon-phonon vertex, and
$g_{\lambda}$ for the electron-phonon vertex as discussed in
the previous section. The bare phonon propagator is given by
\begin{equation}
D_{\lambda}^{0}(\omega)={2\omega_{\lambda}\over{\omega^{2}-
\omega_{\lambda}^{2}+2i\omega_{\lambda}\Gamma_{\lambda}^{i}}},
\end{equation}
where $\Gamma_{\lambda}^{i}$ is the intrinsic phonon linewidth resulting
from e.g. the decay of the phonon caused by the presence of an
anharmonic lattice potential. In the metallic state, the
electron-phonon coupling renormalizes the phonon propagator
such that
\begin{equation}
D_{\lambda}(\omega)={D_{\lambda}^{0}(\omega)\over{1+g_{\lambda}^{2}
D^{0}_{\lambda}(\omega)\chi_{\lambda}(\omega)}},
\end{equation}
where $\chi_{\lambda}$ is the complex electronic susceptibility
evaluated in channel $\lambda$\cite{kubo}.
Defining the renormalized phonon frequency through
\begin{equation}
\hat\omega_{\lambda}^{2}=\omega_{\lambda}^{2}
(1-\lambda(\omega)),
\end{equation}
where
$\lambda(\omega)=2g_{\lambda}^{2}\chi_{\lambda}^{\prime}
(\omega)/\omega_{\lambda}$
and $\chi^{\prime}$ denotes the real part of the susceptibility,
we arrive at the following expression for the renormalized phonon
propagator
\begin{equation}
D_{\lambda}(\omega)={2\omega_{\lambda}\over{\omega^{2}-
\hat\omega_{\lambda}^{2}+
2i\omega_{\lambda}\Gamma_{\lambda}(\omega)}},
\end{equation}
where $\Gamma_{\lambda}$ is the total frequency dependent
linewidth of the phonon,
$\Gamma_{\lambda}(\omega)=
\Gamma_{\lambda}^{i}+g_{\lambda}^{2}\chi_{\lambda}^{\prime\prime}(\omega)$.

We now sum up the diagrams in Fig. (3) for the Raman response in
terms of the susceptibility $\chi_{\lambda}$. After some lengthy but trivial
algebra we obtain
\begin{eqnarray}
\chi_{\lambda,full}^{\prime\prime}(\omega)&=&
{(\omega+\omega_{a})^{2}\over{(\omega^{2}-\hat\omega_{\lambda}^{2})^{2}+
[2\omega_{\lambda}\Gamma_{\lambda}(\omega)]^{2}}}
\Bigg\{\gamma_{\lambda}^{2}\chi_{\lambda}^{\prime\prime}(\omega)\left[
(\omega-\omega_{a})^{2}+4\Gamma_{\lambda}^{i}\Gamma_{\lambda}(\omega)
\left({\omega_{\lambda}\over{\omega+\omega_{a}}}\right)^{2}\right]
\nonumber \\
& &+4g_{p-p}^{2}\Gamma_{\lambda}^{i}
\left({\omega_{\lambda}\over{\omega+\omega_{a}}}\right)^{2}
[1+\lambda(\omega)/\beta]^{2}\Bigg\},
\end{eqnarray}
with
\begin{equation}
\beta={2g_{p-p}g_{\lambda}\over{\gamma_{\lambda} \omega_{\lambda}}},
\ \ \ \ \ \
\gamma_{\lambda}=\langle\phi_{\lambda}({\bf k})\gamma({\bf k})\rangle.
\end{equation}
Here $\langle \cdots \rangle$ denotes an average over the Fermi surface.
Thus $\gamma_{\lambda}$ represents the symmetry elements of the Raman
vertex projected out by the incoming and outgoing
photon polarization vectors.

Eq. (32) is a generalized form of the Breit-Wigner or Fano lineshape
describing the interaction of a discrete excitation (phonon) with an
electronic continuum. Here the frequency $\omega_{a}=
\omega_{\lambda}\sqrt{1+\beta}$ sets the position of
the antiresonance of the lineshape. If the intrinsic phonon width
$\Gamma^{i}_{\lambda}=0$ then at the antiresonance
$\chi^{\prime\prime}_{\lambda,full}(\omega=\omega_a)=0$.
Since $\chi_{\lambda}$ describes the
electronic contribution to Raman scattering, which is generally featureless
and a smooth function of frequency, we can replace everywhere $\chi_{\lambda}$
by the value it takes on near the phonon frequency to fit the Raman
spectra in the vicinity of the phonon. However to fit the entire
spectrum in the normal state, the full $\omega$ dependence is required.

The description of the continuum in high T$_{c}$ superconductors has
been the focus of a large body of both theoretical and experimental work.
While at present no theory can completely describe the full
symmetry-dependent continuum for a large range of frequency shifts, we
now give expressions for the continuum in two separate cases. In the
presence of impurity scattering, the continuum lineshape has a simple
Lorenztian form \cite{zawa2}
\begin{equation}
\chi_{\lambda}^{\prime\prime}(\omega)=N_{F}
{\omega\tau_{\lambda}^{*-1}\over{\omega^{2}+\tau_{\lambda}^{*-2}}}.
\end{equation}
Here $1/\tau_{\lambda}^{*}=1/\tau_{m=0}-1/\tau_{\lambda}$ is the
channel-dependent impurity scattering rate reduced by vertex corrections.
While this form can fit the data on the cuprate materials
at least in the $B_{1g}$ channel (i.e., crossed polarization
orientations aligned 45 degrees with respect to the $a-b$ axis in
the Cu-O planes) at low frequencies, the high frequency
tail which remains relatively constant and the frequency behavior at low
frequencies in other channels cannot be accounted for.

Another form for
the continuum can be arrived at by taking nesting of the Fermi surface into
account as has been done by one of the present authors\cite{viro}. The
Raman susceptibility has a similar form as Eq. (34) with the impurity
scattering rate replaced by the addition of the
electron-electron and impurity scattering rates on a nested Fermi surface
and the effect of mass renormalization taken into account,
\begin{eqnarray}
& &\tau_{\lambda}^{*-1}\rightarrow
\tau_{\lambda}^{*-1}+\alpha\sqrt{(\beta^{\prime} T)^{2}+\omega^{2}},
\nonumber \\
& &\omega \rightarrow \omega m^{*}(\omega)/m, \nonumber \\
& &{\rm with}\ m^{*}(\omega)/m=1+{2\alpha\over{\pi}}
{\rm ln}\left[{\omega_{c}\over{\sqrt{(\beta^{\prime} T)^{2}+\omega^{2}}}}\right],
\end{eqnarray}
where $\alpha, \beta^{\prime}$ and $\omega_{c}$ are constants determined by a
fit to the electronic continuum in the normal state.
This form for the Raman susceptibility provides an adequate
description to the continuum in the normal state of both
YBCO and BSCCO (see Ref. \cite{viro})
and thus we will use this form rather than Eq. (34).

In order to fit the Raman spectra in the normal state, one sees from
Eqs. (32) and (35) that a great deal of parameters are required. A
fitting procedure is now discussed that greatly reduces the number of
free parameters. The first step in the procedure is to determine the
bare phonon parameters $\omega_{\lambda}$ and
$\Gamma_{\lambda}^{i}$, which in principle can simply
be read off from a fit to the lineshape of the phonon in the
insulating state where $g_{\lambda}=0$. While the intrinsic width of the 
phonon can be read off as $\Gamma_{\lambda}^{i}=2.5 $cm$^{-1}$, a problem
arises with the position of the $B_{1g}$ phonon, since it
shows only a small renormalization with doping from the insulator to the
metallic state \cite{reznik}, the origin of which is unknown. 
This could be due in part to changes of
the lattice parameters with doping and/or the tetragonal to orthorhombic
transition\cite{rev}. 
Therefore, we thus must keep this parameter free and use the 
bare frequency we obtain from fitting the normal state. While in 
principle $g_{p-p}$ can
be determined as well at this stage, since the intensity of the spectra
is given in arbitrary units, this parameter (which sets the overall
intensity) must also remain free.

Turning next to the metallic state,
the continuum parameters $\alpha, \ \beta^{\prime},$ and $\gamma_{\lambda}$
are tuned
to fit the full frequency range of the continuum minus the contribution
of the phonons. We then tune the renormalized phonon position,
$\hat\omega_{B_{1g}}$, which we later can compare with the result predicted
from the value of $g_{B_{1g}}$ obtained in
Section II. Thus the only unknown parameters are $g_{p-p}$ (and thus $\beta$),
which also sets the antiresonance position $\omega_{a}$, and the bare
phonon frequency. 

With these constrainsts, such a procedure produces the theory lines shown in
Fig. (4a) compared to the data of Ref. \cite{rudi1} on YBCO.
The parameters used
to obtain the fit are as follows: 
$\omega_{B_{1g}}=358 {\rm cm}^{-1},
\hat\omega_{B_{1g}}=348$ cm$^{-1},
\Gamma_{B_{1g}}=4.9$ cm$^{-1},\Gamma_{B_{1g}}^{i}=2.5$ cm$^{-1}, 
\alpha=0.55, \beta^{\prime}=3.3,
g_{B_{1g}}^{2}N_{F}/\omega_{B_{1g}}=0.0624,
\tau_{B_{1g}}^{*-1}=600$ cm$^{-1}, \omega_{c}=12,000 $cm$^{-1}, $and$
g_{p-p}/\gamma_{B_{1g}}=0.0026$, which sets
$\omega_{a}=360$ cm$^{-1}$. Similar fits have been obtained
before via the usual Fano expression\cite{girsh,rev} albeit with unconstrained
parameters determined solely via a fitting routine.
The value of the
coupling constant used is roughly 20 percent smaller than the one
given by the crude approximation
in Section II.  Given the uncertainty involved in the band parameters
and the neglect of
screening (which will reduce the crystal field and thus lower the coupling
constant), we remark that the
coupling constant derived from a microscopic picture of charge
transfer within the unit cell provides an accurate description of the
strength of the asymmetric Fano lineshape of the $B_{1g}$ phonon in
YBCO.

\section{superconducting state}

Turning now to the superconducting state, we note that Eq. (32) is still
a valid description of the Raman lineshape provided that we use the
Raman susceptibility calculated for the superconducting case. In fact,
it is possible to obtain important information about the symmetry of the
electron pairing in a superconductor through an investigation of the
changes in the phonon lineshape below T$_{c}$. In particular, the
electron-phonon coupling depends crucially on the
symmetries of both the phonon and the order parameter. The role of
the coupling between a phonon of given symmetry and an order parameter of
a different symmetry was recently discussed by one of the authors \cite{tpd},
where it was shown that changes in the phonon lineshape are the greatest
for a phonon which possesses the same symmetry as the order parameter while
smaller changes are predicted for phonons of a symmetry orthogonal to that
of the energy gap. This is pictorially shown in Fig. 2, which shows the
energy gap $\Delta({\bf k})$ squared and the
electron-phonon vertex $g({\bf k})$ squared, which enters into the Raman
response function $\chi$ as a weighted average around the Fermi surface.
Thus the phonon vertex and the energy gap will contructively (destructively)
interfere with each other if they have the same (orthogonal)
symmetry. This fact can be used to determine the predominant energy gap
symmetry.

In Ref.
\cite{Dev} the electronic contribution to Raman scattering
$\chi_{\lambda}$ was calculated in
detail for a superconductor with $d_{x^{2}-y^{2}}$ pairing symmetry
and good fits were obtained to the Raman spectra of BSCCO. For details
of the theory, we direct the reader to that Reference and simply
write down the Raman response function obtained for the $B_{1g}$
channel using a $d_{x^{2}-y^{2}}$ gap,
$\Delta({\bf  \hat k},T)=\Delta_{0}(T)\cos(2\varphi)$ for a cylindrical Fermi
surface (here $x=\omega/\Delta_{0}$):
\begin{equation}
\chi_{B_{1g}}^{\prime\prime}=\tanh(\omega/4T)
{2N_{F}\over{3\pi x}}
\cases{[ (2+x^{2})K(x)-2(1+x^{2})E(x)], & $x\le 1$, \cr
x[ (1+2x^{2})K(1/x)-2(1+x^{2})E(1/x)],
& $x > 1$.}
\end{equation}
The real parts are obtainable through Kramers-Kronig analysis (see
Ref. \cite{tpd}).
The resulting response has a logarithmic divergence at the pair edge
(an artifact of the two-dimensional nature of the Fermi surface) and
vanishes as $x^{3}$ for small frequencies. To make a fit to the
continuum in the superconducting state, we first convolute Eq. (36)
with a Gaussian to mimic the effect of finite $z$ dispersion of the
Fermi surface, experimental resolution, etc. Fitting the
resulting function to the
superconducting state continuum for YBCO leads to value $\Delta=240$ cm$^{-1}$
and a smearing width $\Gamma/\Delta=0.2$ (not to be confused with the
intrinsic phonon linewidth). Then, Eq. (36) can be used to draw a fit
to the full $B_{1g}$ response using the parameters obtained from the
fit to the response in the normal metal.

The resulting fit to the lineshape using Eq. (36) and its corresponding
real part (determined via Kramer's Kronig transformation) at $T=20 K$ is
shown in Fig. (4b) compared to the data on YBCO from Ref. \cite{rudi2}.
Here the parameters used are the same as in the normal
metal case (Section III) with the only change resulting in the use of
Eq. (36) and its real part rather than Eqs. (34-35). Using the coupling
constant $\lambda=0.0624$ as obtained in fitting the normal state leads to
$\hat\omega_{B_{1g}}=337$ cm$^{-1}$ and  $\Gamma_{B_{1g}}=7.9$ cm$^{-1}$,
i.e., the phonon broadens and softens at low temperatures compared to its
normal state lineshape.

The fit points towards the predominance of an energy gap of $B_{1g}$
symmetry for the following reasons. First we remark that the theory
predicts the correct magnitude and sign of the phonon renormalization
due to superconductivity. This would be in marked contrast if the gap
was $s-$wave or of $d-$wave nature but of other symmetry
($d_{xy},d_{3z^{2}-r^{2}}$,etc). In the first case an isotropic gap would
lead to trivial coupling of the energy gap to the symmetry of the phonon
and all phonons would show the same qualitative behavior, which is not
the case for the cuprates \cite{rev}. In the second case, previous
work has demonstrated that the channel probed via Raman scattering which
shows a peak at the highest frequency in the continuum in the superconducting
state of all the channels gives the predominant symmetry of the energy gap.
This again points to $d_{x^{2}-y^{2}}$ pairing. Further, if the gap was
of another symmetry (i.e., $B_{2g}$ or $A_{1g}$) then the position of the
phonon, $\omega_{B_{1g}}/\Delta=1.42$, would predict small phonon
softening since $(i)$ the real part of the phonon self energy
changes from negative to positive values at roughly $\omega/\Delta=1.5$,
and (ii) the weight $\langle \gamma^{2}({\bf k})\mid\Delta({\bf k})\mid^{2}
\rangle$ over the Fermi surface which determines the coupling
would be smaller for a gap of different symmetry than the vertex.
Therefore, while although Raman cannot distinguish between and energy gap
which changes sign or not around the Fermi surface, the evidence from the
data indicate that the electronic pairing is predominantly of
$d_{x^{2}-y^{2}}$ symmetry.

\section{conclusions}

In order to improve the study of the superconducting gap anisotropy in
high temperature superconductors\cite{Dev,tpd} the interactions of the 
electronic continuum and the characteristic phonons shown in Fig. 1 have
been included in the theory. These interactions and the intrinsic width
of the phonons had complicated a previous comparison of the theory for the
continuum for certain gap anisotropy to the experimental
data since the contribution of the phonons had to be subtracted in an
intuitive way (see Ref. 2). In addition, the phonon lineshape is of course
distorted by the Fano interference.

The starting point of our calculation was the model calculation of the 
electron-phonon interaction based on Ref. 6 and a simple tight binding
calculation for the electrons is used. The latter is certainly not very 
adequate but demonstrates the correct characteristic symmetry
properties of the electronic wave functions.

Our main conclusions are as follows: 

(i) {\it Phonon lineshape} - The
asymmetric Fano lineshape of the $B_{1g}$ phonon is reproduced while the
interaction of the continuum with the $A_{1g}$ phonon is much weaker
according to the experiments \cite{rev} and disappears in our model
calculation (see Sec. II B). Using more adequate electronic wave
functions, this coupling may reappear but still must be weaker than in
the case of the $B_{1g}$ phonon; 

(ii) {\it Electric Field} - The perpendicular
electric field acting on the oxygen atoms in the CuO$_{2}$ plane has been
determined by a comparison with the experiments in the normal phase,
assuming tetragonal symmetry, i.e., $\vec E=\vec E_{x}=\vec E_{y}$.
This symmetry is certainly destroyed by the presence of the chains as
shown in Fig. 1, and results in a mixing of the $A_{1g}$ and $B_{1g}$
symmetry channels. It is obvious from the coupling presented in Sec. II B
and from the calculation of the diagrams depicted in Fig. 3, that starting
with, e.g., light polarizations assigned to $B_{1g}$ symmetry, the 
coupling with $\vec E_{x} \ne \vec E_{y}$ leads to a contribution to
the electronic continumm of $A_{1g}$ symmetry which is further coupled 
to the $A_{1g}$ phonons. The $A_{1g}$ phonons do appear in the $B_{1g}$
spectrum observed by experiments (see Fig. 4a), but it is not clear
whether this is dominantly due to misalignments of the samples or to the
channel mixing. A more elaborate calculation of the electric fields
$\vec E_{x}$ and $\vec E_{y}$ arising from different ions would
certainly help to resolve this question; 

(iii) {\it Gap anisotropy} - The
excellent fit of the combined electronic continuum + phonon spectrum in
Fig. 4b is obtained by assuming the simplest form of the superconducting
gap of $d_{x^{2}-y^{2}}$ type for $B_{1g}$ symmetry. The parameters are
borrowed from the fitting of the normal state. The only additional
parameters introduced are the gap amplitude defined above Eq. (36)
and an additional smearing $\Gamma \sim 0.2\Delta_{0}$ with a non-unique
origin (see Sec. IV). This fit certainly supports the previous fit in Ref.
\cite{Dev} where the phonons had been subtracted. 

Concerning the anisotropy of the gap it has already been pointed out
that Raman scattering provides information only concerning the
absolute value of the gap. The fit of the low energy spectra in the 
present paper clearly demonstrates that assuming $d-$wave pairing
in the superconductor, the gap with $d_{x^{2}-y^{2}}$ symmetry gives an
excellent fit but $d_{xy}, d_{xz},$ and $d_{yz}$ are not applicable
(see e.g., Ref. \cite{Dev}). A very highly anisotropic extended $s-$wave
state which has predominantly $B_{1g}$ symmetry cannot be ruled out. 

For
instance recently the photoemission experiment of Ref. 
\cite{argonne} suggests a
gap which can be represented by the form $\Delta(\varphi)=\Delta_{0}
[1+a\cos(4\varphi)]/(1+a)$, with the parameter $a$ just exceeding one.
Our fits clearly show that a fit with $a=1$ are not satisfactory.  The 
reason is that the different contributions of the
different gap features to the low energy 
($\omega<<\Delta_{0}$) part of the spectra with e.g. $B_{1g}$ symmetry:
(i)the large gap around $\varphi={\pi\over{2}}n, (n=0,1,2,\cdots)$ does not
contribute (see Fig. 2); (ii) the gap changing sign at $\varphi={\pi\over{4}}
+n{\pi\over{2}}$ contributes as $\omega^{3}$ at low frequencies; (iii) the
gap touching zero with zero slope (parameter $a=1$) at $\varphi={\pi\over{4}}
+n{\pi\over{2}}$ contributes as $\omega$ at low frequencies
(see Ref. \cite{Dev}); (iv) the gap with zero slope but not at zero energy
(parameter $a \ne 1$) gives a sharp discontinuity and approximately linear 
term (modulo logs) above or below it ($a < 1$ and $a>1$, respectively).

On the basis of the above feautures, the zero slope at energy larger than
$\sim 20$cm$^{-1}$ and at $\varphi={\pi\over{4}}+n{\pi\over{2}}$ can be
ruled out. Any feature with a gap minimum on a smaller energy scale can
however still be possible, but the region of the Fermi surface where the
minima occur must be very narrow compared with those given by
the expression mentioned above. As the minima should occur in an anomalously
narrow region, the slopes coming or leaving the minima must be very large.
Such a behavior appear to us as very unlikely.

In summary, the main result of the present paper is the confirmation of
the conclusion of Refs. \cite{Dev,tpd}, namely, that the Raman spectra are
in excellent agreement with a superconducting gap of $d_{x^{2}-y^{2}}$
symmetry, but gaps with very sharp features on a small part of the 
Fermi surface cannot be ruled out.

\acknowledgments

The authors would like to acknowledge stimulating discussions with
G. T. Zim\'anyi, B. Stadlober, D. Reznik, R. Hackl, J. C. Irwin, D. Einzel,
and M. Cardona. One of the authors (T.P.D.)
would like to acknowledge the hospitality of the Research Institute for
Solid State Physics of the Hungarian Academy of Science and the Institute
of Physics of the Technical University of Budapest
where parts of this work were completed.
This work was supported by the US-Hungarian
Science and Technology Joint Fund under Project No. 265, NSF Grant number
92-06023, and by the
Hungarian National Research Fund under Grants No. OTKA2950, 7283
and T4473.

\figure{The unit cell of the CuO$_2$ plane in YBa$_2$Cu$_3$O$_y$
is shown with the atomic
displacements corresponding to the phonons of $A_{1g}$ (a.) and $B_{1g}$
(b.) symmetry. A finite electric field {\bf E} perpendicular to the
planes due to the asymmetric environment (Y above, Ba below) is indicated.
This induces the respective charge transfer fluctuations ($\delta\rho$)
denoted by the two way arrows.}

\figure{The dashed line above the cylindrical Fermi surface (F.S.) represents
a $d_{x^2-y^2}$ superconducting gap $|\Delta(\phi)|^2\sim \cos^2(2\phi)$.
The areas shaded by the vertical dotted lines mark the regions of the
F.S. which are probed by Raman scattering in the $A_{1g}$ (a.) and in the
$B_{1g}$ (b.) geometries.}

\figure{Feynman diagrams depicting the three contributions to Raman
scattering in metals. Here the solid line is the electron Green's function,
the wavy line the phonon propagator, and the dashed line the photon
propagator. The vertices are defined as in the text.}

\figure{(a).Fit of Eq. (32), (34-35) to the $B_{1g}$ spectra from 
Reference \cite{rudi1} on YBCO at
$T=100 K$. The parameters used to obtain the fit are discussed in the
text;(b) Fit of Eqs. (32) and (36) to the $B_{1g}$ spectra from Reference
\cite{rudi2} on YBCO at $T=20 K$. The parameters used to obtain the fit
are discussed in the text. The additional phonons besides the $340$ cm$^{-1}$
have been subtracted off for clarity (see Fig. 4a).}
\end{document}